\documentstyle[12pt,aasms4]{article}
 
\begin{document}
\title{The true and eccentric anomaly parametrizations of the perturbed Kepler
motion}
\author{L\'aszl\'o \'A. Gergely\dag\ddag, Zolt\'an I. Perj\'es\dag and M\'aty\'as
Vas\'uth\dag}
\affil{\dag\ KFKI Research Institute for Particle and Nuclear\\
Physics, Budapest 114, P.O.Box 49, H-1525 Hungary}
\affil{\ddag\ Laboratoire de Physique Th\'{e}orique,\\
Universit\'{e} Louis Pasteur, 3--5 rue de l'Universit\'{e}, 67084\\
Strasbourg, France}
 
\begin{abstract}
The true and eccentric anomaly parametrizations of the Kepler motion are
generalized to quasiperiodic orbits, by considering perturbations of 
the radial part of the kinetic energy in a form of a series of negative 
powers of the orbital radius. 
A toolchest
of methods for averaging observables as functions of the energy $E$ and
angular momentum $L$ is developed. A broad range of systems governed by the
generic Brumberg force and recent applications in the theory of
gravitational radiation involve integrals of these functions over a period
of motion. These integrals are evaluated by using the residue theorem. In
the course of this work two important questions emerge: (1) When does the
true and eccentric anomaly parameter exist? (2) Under what circumstances and
why are the poles in the origin? The purpose of this paper is to find the
answer to these queries.
\end{abstract}
 

\section{Introduction}
 
In a recent series of papers (\cite{GPV1}-\cite{GPV3}), the present
authors
have derived the cumulative back reaction effects of gravitational radiation
emitted by a binary system of spinning components. The purpose of our work
has been to study the effects of the spins on the secular evolution of the
system under gravitational back reaction, necessary in order to provide
realistic signal templates for the designated gravitational wave
observatories LIGO (\cite{LIGO}), VIRGO (\cite{VIRGO}),
GEO (\cite{GEO}) and TAMA (\cite
{TAMA}). Neutron star-black hole and black hole-black hole
binaries are
designated sources (\cite{Thorne}) for the Laser
Interferometer Space Antenna
(LISA, \cite{LISA},\ \cite{LISA1}). In the course of the computation,
two
parameters of the quasiperiodic orbit have been efficiently utilized for
evaluating integrals of the type
\begin{equation}
\int_{0}^{T}\frac{\omega }{r^{2+n}}dt\ , \label{inttype}
\end{equation}
where $T$ is a radial period and $\omega $ is a function that has no
manifest radial dependence (in a sense that will be made more precise in
Section III). The radial period, for instance, is determined by this
integral with $n=-2$, while the instantaneous radiative losses of the
quantities characterizing the orbit are sums of contributions of type (\ref
{inttype}) with $n=1,..,7.$
 
For averages with $n\geq 0$, a parameter $\chi $ generalizing the true
anomaly of the Kepler motion renders the poles of the integrand in the
origin (after passing to the variable $z=e^{i\chi}$),
thereby easing the evaluation of the integral. For $n<0$, on the
other hand, the parameter $\xi $ generalizing the eccentric anomaly proves
suitable. Hence in both cases the residue theorem can be applied to
integrands with a pole in the origin of the complex plane and, in
certain cases, with an additional pole. This is to be
contrasted with the elaborate structure of the integrands when performing a
radial integration over the same quantities (\cite{Rieth}).
 
Our purpose in this work is to gain an insight into the nature of these two
parametrizations of the perturbed Kepler motion, and to systematically search 
the structure of the integrals of the type (\ref{inttype}). We show that these
parametrizations exist for a wide class of quasiperiodic orbits. Our method
of averaging is independent of the problem of radiation reaction. We
therefore expect that the method developed here will find applications,
beyond the theory of gravitational radiation, in the treatment of
astrophysical binary systems.

We will study perturbations of the Kepler motion represented by a series of 
negative powers of the radial coordinate $r$ in the expression of 
${\dot r}^2$. 
This type of perturbation is generic 
enough to include both the radiation reaction perturbative terms of our 
previous works as well as to comprise other physically relevant situations. 
 
A wide class of perturbed Kepler motions, which fit in our description, 
is described by the Brumberg force
(\cite{Brumberg},\ \cite{Soffel})
\begin{equation}
\ddot{{\bf r}}=-\frac{m{\bf r}}{r^{3}}+\frac{m{\bf r}}{r^{3}}\left[ \frac{%
2\gamma m}{r}-2\beta {\bf \dot{r}}^{2}+3\alpha \frac{({\bf r\dot{r}})^{2}}{%
r^{2}}\right] +{\frac{2\lambda m({\bf r\dot{r}})}{r^{3}}}{\bf \dot{r}}\ ,
\label{brumbg}
\end{equation}
which can be derived from the Lagrangian
\begin{equation}
{\cal L}=\frac{\mu }{2}{\bf \dot{r}^{2}}+\frac{m\mu }{r}+\frac{1}{4}\left(
\alpha -\beta +\frac{\lambda }{2}\right) \mu {\bf \dot{r}^{4}}+\left( \beta
-\alpha +\frac{\lambda }{2}\right) \frac{m\mu }{r}{\bf \dot{r}^{2}}+\left(
\beta -\gamma +\frac{\lambda }{2}\right) \frac{m^{2}\mu }{r^{2}}+\frac{%
\alpha m\mu }{r^{3}}({\bf r\dot{r}})^{2}\ .  \label{brlagr}
\end{equation}
 
Here $\alpha, \beta, \gamma$ and $\lambda$ are perturbation parameters, the
total mass is $m=m_1+m_2$ and the reduced mass is $\mu~=~m_1m_2/m$.
 
The energy $E=\left( \partial {\cal L}/\partial {\bf \dot{r}}\right) {\bf
\dot{r}-}{\cal L}$ and the magnitude $L$ of the angular momentum
\begin{equation}
{\bf L}={\bf L_{N}}\left( 1+\frac{2\alpha -2\beta +\lambda }{\mu }E+\frac{%
2\lambda m}{r}\right)  \label{angmom}
\end{equation}
of the system are conserved. In (\ref{angmom}) ${\bf L_{N}}=\mu {\bf r}%
\times {\bf \dot{r}}$ is the Newtonian angular momentum. By substituting in
the Keplerian part of the energy $E$ the identity
\begin{equation}
{\bf \dot{r}^{2}}=\dot{r}^{2}+\frac{L_{N}^{2}}{\mu ^{2}r^{2}}  \label{v^2}
\end{equation}
(with $L_{N}^{2}$ expressed from (\ref{angmom}) in terms of $L^{2}$) and
inserting the Keplerian expression ${\bf \dot{r}^{2}}=2E/\mu +2m/r$ in the
perturbative terms of $E$, the radial equation of motion is found:
\begin{eqnarray}
\dot{r}^{2}=\frac{2E}{\mu }+\frac{2m}{r}-\frac{L^{2}}{\mu ^{2}r^{2}}
&-&(2\alpha -2\beta +\lambda )\frac{3E^{2}}{\mu ^{2}}-(3\alpha -2\beta
+2\lambda )\frac{4mE}{\mu r}  \label{radialBru} \\
&-&(3\alpha -2\beta +2\lambda +\gamma )\frac{2m^{2}}{r^{2}}+(2\alpha -2\beta
+\lambda )\frac{2EL^{2}}{\mu ^{3}r^{2}}+(\alpha +2\lambda )\frac{2mL^{2}}{%
\mu ^{2}r^{3}}\ .  \nonumber
\end{eqnarray}
 
With $\alpha=\nu/2, \beta=(1+3\nu)/2, \gamma=2+\nu, \lambda=2-\nu$ we get
the radial equation of \cite{DD}.
 
The other relevant example which fits in the generic perturbation scheme 
to be treated in this work is the dynamics of the spinning binary system
(\cite{BB}, \cite{Kidder}). The major difference compared to the 
Brumberg-type perturbation is that the orbital angular momentum ${\bf L}$ 
is not parallel to ${\bf L_{N}.}$ For comparison, we give the radial 
equation characterizing the spinning binary system, up to linear terms 
in spin (\cite{GPV3}):
\begin{equation}
\dot{r}^{2}=\frac{2E}{\mu }+\frac{2m}{r}-\frac{L^{2}}{\mu ^{2}r^{2}}+{\frac{%
2E({\mbox {\boldmath {\bf L}$\cdot \sigma$}})}{m\mu ^{2}r^{2}}}-{\frac{2(2%
{\bf L\cdot S}+{\mbox {\boldmath {\bf L}$\cdot\sigma$}})}{\mu r^{3}}}\ .
\label{radial}
\end{equation}
Here ${\bf S}$ and ${\mbox{\boldmath $\sigma$}}$ are the total and the
weighted spin vectors:
\begin{equation}
{\bf S}={\bf S_{1}+S_{2}}\ ,\qquad \mbox{\boldmath $\sigma$}=\eta {\bf S_{1}}%
+\eta ^{-1}{\bf S_{2}}\ ,\qquad \eta ={\frac{m_{2}}{m_{1}}}\ .
\end{equation}
 
In the next section we study the radial component of a perturbed Keplerian
motion generic enough to include both (\ref{radialBru}) and (\ref{radial}).
We express the turning points of such a radial motion in terms of constants
of the motion.
 
Then in Sec. III, we solve the radial motion by introducing the true anomaly
parametrization and we formulate rigurous statements about the poles of the
integrands (\ref{inttype}), for $n\geq 0$. Integrands with $n<0$ will be
discussed in a similar fashion in Sec. IV, after introducing the eccentric
anomaly parametrization. An immediate application of the developed methods
is the computation of the radial period for a wide class of perturbed
Keplerian motions. This is done in Sec. V. Finally in Sec. VI. we summarize
our results. Throughout the paper we use relativistic units $G=c=1$.
 
\section{The radial motion}
 
We consider the radial equation of a perturbed Kepler problem\footnote{
For a review of the Newtonian and relativistic perturbed Kepler problem,
{\sl cf.} \cite{DD}.}
\begin{equation}
\dot{r}^{2}=\frac{2E}{\mu }+\frac{2m}{r}-\frac{L^{2}}{\mu ^{2}r^{2}}%
+\sum_{i=0}^{p}\frac{\varphi _{i}}{\mu ^{2}r^{i}} \ , \label{rdot}
\end{equation}
where the small coefficients $\varphi _{i}$ couple the perturbations to the
system. The first three terms on the right are familiar from the radial
Kepler motion, however, the constants $E$ and $L$ characterize the perturbed
motion. In a Lagrangean treatment, the conservation of the energy $E$ and
the total angular momentum ${\bf J}$ follows from symmetry considerations.
We assume these symmetries to characterize the system to the required order.
Furthermore, we assume that the total spin ${\bf S}$ of the system evolves
in a direction orthogonal to the orbital momentum ${\bf L}$, a requirement
which is met by the spinning binary system and has the immediate consequence
that $L$ is conserved. In what follows, we shall employ a zero subscript for
denoting the parts of expressions taken from the Kepler motion. For example,
\begin{equation}
\dot{r}_{0}^{2}=\frac{2E}{\mu }+\frac{2m}{r}-\frac{L^{2}}{\mu ^{2}r^{2}}\ .
\label{rdot0}
\end{equation}
 
The turning points at $\dot{r}^{2}=0$ are the solutions of the algebraic
equation:
\begin{equation}  \label{turning}
2\mu Er^{p}+2m\mu ^{2}r^{p-1}-L^{2}r^{p-2}+\sum_{i=0}^{p}\varphi
_{i}r^{p-i}=0.\
\end{equation}
 
Without the perturbing terms this Eq. has the roots: $r_{+}$, $r_{-}$,
0,..., 0, where there are $p-2$ zero roots,
\begin{equation}
r_{\pm }=\frac{m\mu \pm A_{0}}{-2E}\ =\frac{L^2}{\mu(m\mu \mp
A_{0})}\qquad \end{equation}
and $A_{0}$ is the length of the Runge-Lenz vector, given by
 
\begin{equation}
\mu A_{0}^{2}=m^{2}\mu ^{3}+2EL^{2}\ .  \label{A0}
\end{equation}
 
Then the roots of Eq. (\ref{turning}) describing the perturbed motion can be
found in the form: $r_{\max}=r_{+}+\varepsilon _{+}$, $r_{\min}=r_{-}+%
\varepsilon _{-}$ and (for $p>2$) $\varepsilon _{k}$, $k=1,...,p-2$,
where $\varepsilon _{k}$ is real or imaginary:
 
\begin{equation}
\varepsilon _k=\left( \frac{\varphi _p}{L^2}\right) ^{1/(p-2)}
\end{equation}
and
\begin{equation}
\varepsilon _{\pm }=-\left[ 2\mu Epr_{\pm }^2+2m\mu ^2(p-1)r_{\pm
}-(p-2)L^2\right] ^{-1}\sum_{i=0}^p\varphi _ir_{\pm }^{3-i} \ .
\end{equation}
The roots $\varepsilon _k$, even if real, are unphysical as they are smaller
than the radius $r$. The exponent $1/(p-2)$ renders this approximation
increasingly corrupt as $p\to \infty $.
 
Thus the turning points are:
\begin{equation}
r_{{}_{{}_{\min}^{\max}}}=\frac{m\mu \pm A_0}{-2E}\pm \frac 1{2\mu A_0}%
\sum_{i=0}^p\varphi _i\left[ \frac{\mu (m\mu \mp A_0)}{L^2}\right]^{i-2}\ .
\label{rmima}
\end{equation}
We will need also their reciprocal:
\begin{equation}  \label{rmm}
\frac 1{r_{{}_{{}_{\min}^{\max}}}}=\frac{\mu (m\mu \mp A_0)}{L^2}\mp \frac 1{%
2\mu A_0}\sum_{i=0}^p\varphi _i\left[ \frac{\mu (m\mu \mp A_0)}{L^2}\right]
^{i}\ .
\end{equation}
Note that $A_0$ is defined by Eq. (\ref{A0}), with the constants $E$ and $L$
characterizing the perturbed motion.
 
\section{True anomaly parametrization}
 
The true anomaly parameter $\chi $ of the Kepler ellipse $r=p \left( 1+e\cos
\chi \right) ^{-1}$ lying in the $z $ plane with {\it semilatus rectum} $p=
\frac{L^{2}}{m\mu ^{2}}$ and eccentricity $e=\frac{A_{0}}{m\mu }$ is the
azimuthal angle. A like parametrization can be introduced for the perturbed
motion, but this can be done in several inequivalent ways. An example of
this parametrization has been discussed by \cite{Ryan} for the radial
motion of a test particle orbiting a spinning point mass. When one attempts
to utilize his parameter for averaging the radiative losses, the integrals
involved will contain poles at various location in the complex $\chi $
plane; and this feature is not amenable for doing the integration. However,
adding any term proportional to $\sin \chi $ will not alter the turning
points in $r$. One may take advantage of this freedom in fixing the
generalized true anomaly parameter in such a way that two main features, one
relating the turning points with the scale of the parameter and the other
the simple periodic dependence of the radius on the parameter, are
stipulated for the perturbed motion:
 
\begin{eqnarray}
(a) &{}&\quad r_{\min}=r(0)  \nonumber \\
&{}&\quad r_{\max}=r(\pi )  \label{parchiab} \\
(b) &{}&\quad \frac{dr}{d(\cos \chi )}=-\gamma r^{2}\rightarrow \frac{1}{r}%
=\gamma \cos \chi +\delta \ .  \nonumber
\end{eqnarray}
These two properties are equivalent with:
 
\begin{equation}
\frac{2}{r}=\frac{1+\cos \chi }{r_{\min}}+\frac{1-\cos \chi }{r_{\max}}
\label{parchi}
\end{equation}
The true anomaly parametrization has a remarkable feature, which can be
encapsulated in
 
{\bf Theorem 1}: {\sl The integral }$\int \frac{1}{r^{2+n}}dt${\sl \ , }$%
n\geq 0${\sl \ is given by the residue in the origin of the complex true
anomaly parameter plane }$z=e^{i\chi }${\sl . }
 
In order to establish this theorem, first we pass in the integral to the
true anomaly parameter:
 
\begin{equation}
\int \frac{1}{r^{2+n}}dt=\int \frac{1}{r^{2+n}}\frac{dt}{dr}\frac{dr}{d\chi }%
d\chi =\frac{1}{2}\left( \frac{1}{r_{\min }}-\frac{1}{r_{\max }}\right) \int
\frac{1}{r^{n}}\frac{\sin \chi }{\stackrel{.}{r}}d\chi .  \label{1}
\end{equation}
 
Here we have used (\ref{parchiab}). The last integrand is a product of two
factors. The parametric form of the first one, $1/r^{n}$ is given by the
binomial expansion
\begin{equation}
\left( \frac{2}{r}\right) ^{n}=\sum_{i=0}^{n}C_{n}^{i}\left( \frac{1+\cos
\chi }{r_{\min}}\right) ^{i}\left( \frac{1-\cos \chi }{r_{\max}}\right) ^{n-i}
\ .  \label{2rn}
\end{equation}
Hence this factor is a polynomial in $\cos\chi$.
 
Next we will show that the second factor $\sin \chi /\dot{r}$ is also
regular. For this purpose we write the radial equation of motion (\ref{rdot}
) in terms of the parameter $\chi $. Using (\ref{2rn}), the Kepler part (\ref
{rdot0}) of $\dot{r}^{2}$ becomes:
\begin{equation}
\dot{r}_{0}^{2}=\frac{A_{0}^{2}}{L^{2}}-\frac{L^{2}}{4\mu ^{2}}\left[ \frac{1%
}{r_{\min}}+\frac{1}{r_{\max}}-\frac{2m\mu ^{2}}{L^{2}}+\left( \frac{1}{r_{\min}%
}-\frac{1}{r_{\max}}\right) \cos \chi \right] ^{2}.
\end{equation}
We insert here the expressions (\ref{rmm}) of the turning points
$r_{\min}$ and
$r_{\max}$ in terms of the constants of the motion and introduce the
notation
\begin{equation}
\Delta _{\pm }^{j}=(m\mu +A_{0})^{j}\pm (m\mu -A_{0})^{j}\ .\label{Delta}
\end{equation}
Keeping terms in $\dot{r}_{0}^{2}$ to linear order in $\varphi _{i}$, we
get
\begin{eqnarray}
\dot{r}_{0}^{2}=\frac{A_{0}^{2}}{L^{2}}\sin^2\chi
-\frac{1}{2\mu ^{2}}\sum_{i=0}^{p}\varphi
_{i}\left(\frac{\mu}{L^2}\right)^i\left[
\Delta _{-}^{i} + \Delta _{+}^{i}\cos\chi
 \right] \cos \chi \ .
\end{eqnarray}
Adding the first-order terms present in
the radial Eq. (\ref{rdot}) we finally obtain
\begin{eqnarray}                     \label{rdotchi}
\dot{r}^{2} &=&\frac{A_{0}^{2}}{L^{2}}\sin ^{2}\chi -\sum_{i=0}^{p}\frac{%
\varphi _{i}}{\mu ^{2}}\left( \frac{\mu }{L^{2}}\right)^{i}K_{i}(\chi ),
\end{eqnarray}
where we denote
\begin{eqnarray}
K_{i}(\chi ) &=&\frac{\Delta _{-}^{i}\cos \chi +\Delta _{+}^{i}\cos ^{2}\chi
}{2}-\left( m\mu +A_{0}\cos \chi \right) ^{i}.
\end{eqnarray}
 
A power series expansion in the small coefficients $\varphi _{i}$ leads to
\begin{equation}
\frac{\sin \chi }{\dot{r}}=\frac{L}{A_{0}}+\frac{L^{3}}{2A_{0}^{3}\sin
^{2}\chi }\sum_{i=0}^{p}\frac{\varphi _{i}}{\mu ^{2}}\left( \frac{\mu }{L^{2}%
}\right)^{i}K_{i}(\chi ).  \label{1dotr}
\end{equation}
We will show, that the numerator of the second term is always proportional
to $\sin ^{2}\chi .$ Indeed
 
\begin{equation}
K_{i}(\chi ) =\sum_{k=0}^{i}C_{i}^{k}(m\mu )^{i-k}A_{0}^{k}\Xi _{k}(\chi )\ ,
\end{equation}
where we denoted
\begin{equation}
\Xi _{k}(\chi )=\frac{1-(-1)^{k}}{2}\cos \chi +\frac{1+(-1)^{k}}{2}\cos
^{2}\chi -\cos ^{k}\chi .
\end{equation}
For $k=0$, a $\sin ^{2}\chi $ factor can be taken out. We then find that
this property continues to hold for all even and odd values:
\begin{eqnarray}
\Xi _{0} &=&-\sin ^{2}\chi ,\qquad \Xi _{1}=\Xi _{2}=0\ ,\nonumber \\
\Xi _{2(\ell +1)} &=&\cos ^{2}\chi -\cos ^{2+2\ell }\chi =\sin ^{2}\chi
\sum_{n=0}^{\ell -1}\cos ^{2(n+1)}\chi \ ,\\
\Xi _{2\ell +1} &=&\cos \chi -\cos ^{1+2\ell }\chi =\sin ^{2}\chi
\sum_{n=0}^{\ell -1}\cos ^{2n+1}\chi\ .
\end{eqnarray}
 
Therefore no $\chi $ dependence remains in the denominator of the integrand (%
\ref{1dotr}). We have thus proved that $\sin \chi /\dot{r}$ is regular.
 
When passing to the complex parameter
\begin{equation}
z=e^{i\chi },  \label{z}
\end{equation}
the denominator of the integrand (a polynomial in $\sin \chi $ and $\cos
\chi $) will be an integer power of $z$, leaving the origin as the only
location of the pole.
 
{\em Theorem 1} being proven, we can now specify which are the functions $%
\omega $ in (\ref{inttype}), which will not destroy the advantageous
property of the integrand. As is transparent from the above proof, such a
function {\em can be any polynomial of }$\sin \chi ${\em \ and }$\cos \chi $%
. In particular it can have the form
\begin{equation}
\omega =\omega _{0}+\varphi (\psi ,\dot{r})\ ,  \label{om}
\end{equation}
where $\omega _{0}$ is some constant, and the perturbation term $\varphi
(\psi ,\dot{r})$ is a polynomial in $\sin \psi ,\ \cos \psi $ and $\dot{r}$.
As the azimuthal angle $\psi $ and  the rate of change $\dot{r}$ of the
radius are needed here to Keplerian order:
\begin{equation}
\psi =\psi _{0}+\chi \ ,\qquad \dot{r}={\frac{A_{0}}{L}}\sin \chi ,
\label{psidotr}
\end{equation}
$\ $the function $\omega $ is a polynomial in $\sin \chi $ and $\cos \chi .$
 
The results of this section can be summarized as follows. {\em For the
perturbed Kepler motions }(\ref{rdot}),{\em \ any integral }(\ref{inttype})
{\em with }$n\geq 0${\sl \ }{\em \ satisfying }(\ref{om}), {\em by use of }(%
\ref{parchi}){\em , }(\ref{z}){\em \ and }(\ref{psidotr}){\em \ can be
turned into a complex integral in the variable} $z$.{\em This integral is
simple to evaluate with the residue }${\cal R}${\em \ lying in the origin:}
 
\begin{equation}
\int_{0}^{T}\frac{\omega _{0}+\varphi (\psi ,\dot{r})}{r^{2+n}}dt\ =2\pi i\
{\cal R}\left( z=0\right) \ . \label{result1}
\end{equation}
 
\section{The eccentric anomaly parametrization}
 
Another convenient parametrization of the Kepler ellipse $r=a(1-e\cos \xi )
$ with semimajor axis $a=\frac{p}{1-e^{2}}=\frac{m\mu }{-2E}$ is given by
the eccentric anomaly parameter $\xi $. In terms of the turning points, the
eccentric anomaly parametrization may be defined as follows,
\begin{equation}
{2}{r}=(1+\cos \xi ){r_{\min}}+(1-\cos \xi ){r_{\max}}\ .  \label{parxi}
\end{equation}
Hence
\begin{equation}
{2}\frac{dr}{d\xi }=(r_{\max}-r_{\min})\sin \xi  \ .\label{rxi}
\end{equation}
 
For a perturbed Kepler motion (\ref{rdot}), we use (\ref{parxi}) as the
definition of the generalized eccentric anomaly parametrization. This
parametrization has a remarkable property the main consequence of which is
contained in the following
 
{\bf Theorem 2}: {\sl The integral }$\int \frac{1}{r^{2+n}}dt$ with $n<0$%
{\sl \ is given by the sum of the residues in the origin of the complex
parameter plane }$w=e^{i\xi }$ and [for perturbative terms with negative
powers $i\ge 2-n$ of $r$ in the radial equation (\ref{rdot})] at
\begin{equation}
w_{1}=\left( \frac{m\mu ^{2}-\sqrt{-2\mu EL^{2}}}{m\mu ^{2}+\sqrt{-2\mu
EL^{2}}}\right) ^{1/2}\ .  \label{extrapole}
\end{equation}
{\sl .}
 
We prove this theorem by passing to the true anomaly parameter in the
integral. For convenience we introduce $n^{\prime}=-1-n\geq 0$.
 
\begin{equation}
\int \frac{1}{r^{2+n}}dt=\int r^{n^{\prime}-1}\frac{dt}{dr}\frac{dr}{d\xi }%
d\xi =\frac{r_{\max }-r_{\min }}{2}\int r^{n^{\prime}}\frac{\sin \xi }{r\dot{%
r}}d\xi .  \label{2}
\end{equation}
 
In the last equality we have used (\ref{rxi}). We insert the binomial
expansion of (\ref{parxi}) in the first factor of the last integrand. In
what follows, we will show that the second factor $\sin \xi /r\dot{r}$ is
regular.
 
Multiplying the radial equation (\ref{rdot}) by $r^{2}$ we get
\begin{equation}
(r\dot{r})^{2}=\left( \frac{2E}{\mu }r^{2}+{2m}r-\frac{L^{2}}{\mu ^{2}}%
\right) +\sum_{i=0}^{p}\frac{\varphi _{i}}{\mu ^{2}r^{i-2}}\ .  \label{rrdot}
\end{equation}
Hence, using again Eq. (\ref{parxi}),
we get the equation of the associated Kepler motion in terms of the
parameter $\xi $:
\begin{equation}
(r\dot{r})_{0}^{2}=\frac{E}{2\mu}\left[(r_{\min}-r_{\max})\cos\xi
+r_{\min}+r_{\max}\right]^2+m(r_{\min}-r_{\max})\cos \xi
+m(r_{\min}+r_{\max}) - \frac{L^2}{\mu^2}
\ . \label{rrdot0} \end{equation}
By simple algebra, we can write this as
\begin{equation}
(r\dot{r})_{0}^{2}=\frac{A_{0}^{2}}{-2\mu E}+\frac{E}{2\mu }\left[
r_{\min}+r_{\max}+\frac{m\mu }{E}+(r_{\min}-r_{\max})\cos \xi \right] ^{2}
\ . \label{rrdot0a}
\end{equation}
By use of the quantities $\Delta _{\pm }^{j}$ introduced in (\ref{Delta}),
and following the steps in the derivation of Eq. (\ref{rdotchi}),
the full radial equation (\ref{rrdot}) is written as
\begin{eqnarray}
(r\dot{r})^{2} &=&\frac{A_{0}^{2}}{-2\mu E}\sin ^{2}\xi -\sum_{i=0}^{p}\frac{
\varphi _{i}}{\mu ^{2}}\left[ \frac{\mu }{L^{2}(m\mu -A_{0}\cos \xi )}
\right] ^{i-2}H_{i-2}(\xi ) \ , \nonumber \\
H_{j}(\xi ) &=&(m\mu -A_{0}\cos \xi )^{j}\frac{\Delta _{-}^{j}\cos \xi
+\Delta _{+}^{j}\cos ^{2}\xi }{2}-\left( \frac{-2EL^{2}}{\mu }\right)^{j}\ . \label{rdotxi}
\end{eqnarray}
In the denominator, the quantity $m\mu -A_{0}\cos \xi $ is strictly positive
as can be seen by either substituting $A_{0}$ or comparing with the
Newtonian value of (\ref{parxi}). By a power series expansion we get
\begin{equation}       \label{sinxiperr}
\frac{\sin \xi }{r\dot{r}}=\frac{(-2\mu E)^{1/2}}{A_{0}}+\frac{(-2\mu
E)^{3/2}}{2A_{0}^{3}\sin ^{2}\xi }\sum_{i=0}^{p}\frac{\varphi _{i}}{\mu ^{2}}
\left[ \frac{\mu }{L^{2}(m\mu -A_{0}\cos \xi )}\right] ^{i-2}H_{i-2}(\xi ).
\end{equation}
Again, we can show that the numerator of the second term is proportional to $%
\sin ^{2}\xi $. By using (\ref{A0}), the second term of the function $%
H_{j}(\xi )$ can be written as $(m\mu -A_{0})^{j}(m\mu +A_{0})^{j}$. For $%
j\geq 0$, we write $H_{j}(\xi )$ as an algebraic sum of products of binomial
expansions:
\begin{equation}
H_{j}(\xi )=\sum_{k,\ell =0}^{j}C_{j}^{k}C_{j}^{\ell }(m\mu )^{2j-k-\ell
}A_{0}^{k+\ell }(-1)^{\ell }\left[ \frac{1-(-1)^{k}}{2}\cos ^{\ell +1}\xi +%
\frac{1+(-1)^{k}}{2}\cos ^{\ell +2}\xi -1\right] \ .  \label{K}
\end{equation}
When $k$ and $\ell $ are of like parity, the square bracket in (\ref{K}) has
the form $\cos ^{2q}\xi -1$ hence we can take out a $\sin ^{2}\xi $ factor.
When $k$ is odd but $\ell $ is even or conversely, the sum of pairs of terms
($k,\ell $) and ($\ell ,k$) is proportional to $\cos ^{2q_{1}-1}\xi
-cos^{2q_{2}-1}\xi ,$ with some integers $q_{1}$ and $q_{2}$. This, again,
contains the factor $\sin ^{2}\xi $. Similarly, the corresponding
expressions of $H_{j}(\xi )$ for $j=-1$ and $j=-2$, given in Table 1, are
proportional to $\sin ^{2}\xi $.
 
Thus we have shown that $\sin \xi /r\dot{r}$ is regular.
 
The integrand in (\ref{2}), when written in terms of the the complex
parameter
\begin{equation}
w=e^{i\xi }\ ,  \label{w}
\end{equation}
will contain a polynomial in $w$ in the denominator. The possible roots
lying inside the unit circle are at $w=0$ and at the root of $m\mu
-A_{0}\cos \xi $ given in (\ref{extrapole}). Let us remark that the factor $%
m\mu -A_{0}\cos \xi $ appears in the perturbative terms of Eq. (\ref{2}) on
the power $n^{\prime }+2-i=1-n-i$, which is {\em negative }only for $i\ge
2-n.$ For these values of $i$ the residue at $w_{1}$ is present.
 
The function $\omega $ in (\ref{inttype}) this time {\em can be a fraction
with the numerator being a polynomial of }$\sin \xi ${\em \ and }$\cos \xi $
{\em and the denominator an arbitrary power }$q\geq 0$ ${\em of}${\em \ }$%
m\mu -A_{0}\cos \xi $. In particular it may have the form (\ref{om}). The
expressions for the azimuthal angle $\psi $ and the rate of change of the
radius $\dot{r},$ to be inserted in the perturbative term, are given by:
\begin{equation}
\cos \left( \psi -\psi _{0}\right) =\frac{m\mu \cos \xi -A_{0}}{m\mu
-A_{0}\cos \xi },\qquad \dot{r}=\left( \frac{-2E}{\mu }\right) ^{1/2}\frac{%
A_{0}\sin \xi }{m\mu -A_{0}\cos \xi }.  \label{psidotr2}
\end{equation}
$\ $In the presence of  $\omega $ the residue at $w_{1}$ appears if any of
the conditions $i\ge 2-n$ , $q\geq -n$ are satisfied$.$
 
In summary, {\em for the perturbed Keplerian motion }(\ref{rdot}),{\em \ any
integral }(\ref{inttype}) {\em with }$n<0${\sl \ }{\em \ satisfying }(\ref
{om}), {\em by use of }(\ref{parxi}){\em , }(\ref{w}){\em \ and }(\ref
{psidotr2}){\em \ can be turned into a complex integral in the variable} $w$%
. {\em Evaluation is simple since the residues }${\cal R}${\em \ are in the
origin and }$w_{1}$ ({\em given in} (\ref{extrapole})){\em :}
 
\begin{equation}
\int_{0}^{T}\frac{\omega _{0}+\varphi (\psi ,\dot{r})}{r^{2+n}}dt\ =2\pi i\
\left[ {\cal R}\left( w=0\right) +{\cal R}\left( w=w_{1}\right) \right] .
\label{result2}
\end{equation}
 
If the radial equation (\ref{rdot}) is such that $p<2-n$ and $\varphi $ such
that $q<-n$, then solely the first residue gives a contribution. In other
cases, when the integrand $f(w)$ has also a pole of order $s$ at
$w=w_{1}\neq\infty $ its residue can be conveniently computed by:
\begin{equation}
{\cal R}\left( w=w_{1}\right) =\frac{1}{(s-1)!}\lim_{w\rightarrow w_{1}}
\frac{d^{s-1}}{dw^{s-1}}[(w-w_{1})^{s}f(w)].  \label{residue_formula}
\end{equation}
 
\section{Computation of the period}
 
As a simple application of the method described in the previous section we
compute the radial period for various perturbed Keplerian motions.
 
The unperturbed system ($\varphi _{i}=0$) has the Kepler period
\begin{equation}
T_{K}=2\pi m\left( \frac{\mu }{-2E}\right) ^{3/2}\ .
\end{equation}
The period $T=\int_{0}^{2\pi }\frac{dt}{d\xi }d\xi $
of the perturbed system is the sum of contributions
\begin{equation}
T=T_{K}\left(1 + \sum_{i=0}^{p}T_{i}\right)\ .
\end{equation}
Here the corrections $T_i$ from $\varphi_{i}$
may be obtained by putting $n=-2$ in Eq. (\ref{2}),
substituting $\sin\xi/r\dot r$ from (\ref{sinxiperr}) and eliminating
$r$ by use of (\ref{parxi}). Next, expressing
 $r_{{}_{{}_{\min}^{\max}}}$ from (\ref{rmima}), we get:
\[
T_{i}=-\frac{E}{m\mu ^{2}A_{0}^{2}}\varphi
_{i}\left( \frac{\mu }{L^{2}}\right) ^{i-2}\left[ \frac{m\mu \Delta
_{+}^{i-2}-A_{0}\Delta _{-}^{i-2}}{2}+\frac{1}{2\pi }\int_{0}^{2\pi }\frac{
H_{i-2}(\xi )d\xi }{\sin ^{2}\xi (m\mu - A_{0}\cos \xi )^{i-3}}\right]
 \ .
\]
 
By using the values of the functions $\Delta _{\pm }^{i}$ and $H_{i}(\xi )$
given in Table 1, we find that the correction $T_{i}$ to
the period vanishes for $i=2,3$. In these
cases, the poles of the period are in the origin of the complex $w$ plane.
For $i=4$, the period is altered by a perturbative term given in the last
column of the Table.
 
 \cite{DD} compute the period $T$ for perturbations of
the form $\varphi _{0}=3E^{2}(3\mu-m )/mc^{2}$ and $\varphi
_{1}=2GE(7\mu-6m )\mu /c^{2}$. (Damour and Deruelle enlist the
constants $A,...,H$ in an alphabetic
notation; thus our energy $E$ is distinct from their constant $E$).
Their result can be recovered by summing up
the corrections to $T_{K}$ for the entries $i=0$ and $i=1$ in our Table 1
for $c=1$.\medskip \mathstrut \bigskip
 
\begin{table}[tbp]
\caption{Values of the constants $\Delta_{\pm}^i$, the function $H_i(\xi)$
and corrections to the period $T$}
\begin{tabular}{lrrrc}
$i$ & $\Delta^{i-2}_+$ & $\Delta^{i-2}_-$ & $H_{i-2}(\xi)$ &
Correction $T_{i}$
 \\
\tableline 0 & $\frac{\mu}{E^2L^4}(m^2\mu^3+EL^2)$ & $-\frac{m\mu^3A_0}{
E^2L^4}$ & $-\frac{m^2\mu^4}{4E^2L^4}\frac{\sin^2\xi}{(m\mu - A_0\cos\xi)^2}$
& $-\frac34\frac{\varphi_0}{\mu E}$ \\
1 & $-\frac{m\mu^2}{EL^2}$ & $\frac{\mu A_0}{EL^2}$ & $\frac{m\mu^2}{2EL^2}
\frac{\sin^2\xi}{m\mu - A_0\cos\xi}$ & $ \frac{\varphi_1}{2m\mu^2}
$ \\
2 & $2$ & $0$ & $-\sin^2\xi$ & $0$ \\
3 & $2m\mu$ & $2A_0$ & $\sin^2\xi(A_0^2-m^2\mu^2+m\mu A_0\cos\xi)$ & $0$
\\
4 & $2(A_0^2\!+\!m^2\mu^2)$ & $4m\mu A_0$ & $\sin^2\xi[ 2m^3\mu^3A_0\cos\xi
\!$ & $
-\frac{E\varphi_4}{mL^4} (m^2\mu^2\!-\!A_0^2)^{1/2}$
\\
  &  & & $
\!-\!A_0^2(A_0^2\!+\!m^2\mu^2)\cos^2\xi \!-\!(A_0^2\!-\!m^2\mu^2)^2]$ &
\end{tabular}
\end{table}
 
\section{Concluding Remarks}
 
We have analyzed in detail two types of generalized Keplerian
parametrizations for which the evaluation of various integrals turns out to
be a simple application of the residue theorem, Eqs. (\ref{result1}) and (
\ref{result2}) encapsulating the main results of this paper. We can state
the generic result, that {\em for a perturbed Keplerian motion }(\ref{rdot})
{\em characterized by some value }$p\geq 0,$ the {\em integrals }(\ref
{inttype}){\em \ with }$n\leq \min (1-p,-1-q)\ ${\em and}$\ n\geq 0${\em ,
when parametrized by the eccentric and true anomaly parameters,
respectively, are given by the corresponding residues at the origin} ${\cal R
}(0).$ For the evaluation of the remaining integrals with $\min
(1-p,-1-q)<n<0$ , (such integrals can appear in the $p\geq 3$ case
irrespective of the value of $q$), the computation of a second residue is
necessary. We have done this in the eccentric parametrization, but a similar
result holds for the true anomaly, where the second pole $z_{1}=-w_{1}$ is
the root of the equation $m\mu +A_{0}\cos \chi =0$.
 
The solution of the radial motion we have presented has certain advantages
over previous approaches (\cite{Rieth},\ \cite{Ryan}) and in our belief
is at
least a viable alternative to the use of the Laplace second integrals for
the Legendre polynomials (\cite{Watson,GI}). Although it is not a
complete parametrization of the motion [a review and comparison of such
parametrizations (\cite{Brumberg,DD},\ \cite{Blandford}-\cite{DD2}) was
given by
\cite{Klioner}], it is well suited when purely radial
effects are considered.
 
The power of the method we have described here was already tested in the
evaluation of integrands involved in the radiative loss computations in \cite
{GPV1}-\cite{GPV3}. All integrands encountered there belong to the class $
n\geq 0$, parametrizable by the true anomaly parameter, with the only pole
in the origin. In this paper we have applied the eccentric anomaly
parametrization for computing the period pertinent to a wide class of
perturbations. A property holding for the spinning binary system, namely $
T=T_{K}$ was shown to be a consequence of the particular type of perturbing
potential governing the motion, rather than a generic feature.
 
\section*{Acknowledgments}
 
This research has been supported by OTKA grants T17176 and D23744. One of
the authors, L.\'A.G. acknowledges the continued support of the Hungarian
State E\"{o}tv\"{o}s Fellowship and of the Soros Foundation.

\end{document}